\documentstyle[11pt]{article}
\makeatletter \@addtoreset{equation}{section} \makeatother

\newcommand{\noi}{\vspace{12pt}\noindent}
\newcommand{\beq}{\begin{equation}}
\newcommand{\eeq}{\end{equation}}
\newcommand{\bea}{\begin{eqnarray}}
\newcommand{\eea}{\end{eqnarray}}

\newcommand{\e}[1]{{(\ref{#1})}}
\newcommand{\eq}[1]{{eq.\ (\ref{#1})}}
\newcommand{\es}[2]{{(\ref{#1}) and (\ref{#2})}}
\newcommand{\eqs}[2]{{eqs.\ (\ref{#1}) and (\ref{#2})}}
\newcommand{\Ref}[1]{{Ref.~\cite{#1}}}
\newcommand{\mb}[1]{{\mbox{${#1}$}}}

\newcommand{\ie}{{${ i.e.\ }$}}

\newcommand{\cf}{{cf.\ }}

\newcommand{\eps}{\varepsilon^{}}

\newcommand{\Deltaone}{\Delta_{1}^{}}
\newcommand{\DeltaE}{\Delta_{E}^{}}

\newcommand{\Deltarho}{\Delta_{\rho}^{}}

\newcommand{\nurho}{\nu_{\rho}^{}}

\newcommand{\str}{{\rm str}}

\newcommand{\Hf}{{1 \over 2}}

\newcommand{\Ih}{{i \over \hbar}}

\newcommand{\deder}[1]{{ 
 {\stackrel{\raise.2ex\hbox{$\leftarrow$}}{\delta^{r}}   } 
\over {   \delta {#1}}  }}
\newcommand{\dedel}[1]{{ 
 {\stackrel{\lower.3ex \hbox{$\rightarrow$}}{\delta^{\ell}}   }
 \over {   \delta {#1}}  }}
\newcommand{\papar}[1]{{ 
 {\stackrel{\raise.2ex\hbox{$\leftarrow$}}{\partial^{r}}   } 
\over {   \partial {#1}}  }}
\newcommand{\papal}[1]{{ 
 {\stackrel{\lower.3ex \hbox{$\rightarrow$}}{\partial^{\ell}}   }
 \over {   \partial {#1}}  }}
\newcommand{\rpa}[1]{{ 
 \stackrel{\raise.2ex\hbox{$\leftarrow$}}{\partial^{r}_{#1}}   }}
\newcommand{\lpa}[1]{{ 
 \stackrel{\lower.3ex\hbox{$\rightarrow$}}{\partial^{\ell}_{#1}}  }}

\newcommand{\proofbox}{\begin{flushright}
${\,\lower0.9pt\vbox{\hrule \hbox{\vrule
height 0.2 cm \hskip 0.2 cm \vrule height 0.2 cm}\hrule}\,}$
\end{flushright}}

\begin{document}
\thispagestyle{empty}
\title{\Large{\bf Odd Scalar Curvature in Anti-Poisson Geometry}}
\author{{\sc Igor~A.~Batalin}$^{1}$ \\I.E.~Tamm Theory Division\\
P.N.~Lebedev Physics Institute\\Russian Academy of Sciences\\
53 Leninsky Prospect\\Moscow 119991\\Russia\\~\\and\\~\\
{\sc Klaus~Bering}$^{2}$\\Institute for Theoretical Physics \& Astrophysics\\
Masaryk University\\Kotl\'a\v{r}sk\'a 2\\CZ-611 37 Brno\\Czech Republic}
\maketitle
\vfill
\begin{abstract}
Recent works have revealed that the recipe for field-antifield quantization
of Lagrangian gauge theories can be considerably relaxed when it comes to
choosing a path integral measure $\rho$ if a zero-order term $\nu_{\rho}$
is added to the $\Delta$ operator. The effects of this odd scalar term
$\nu_{\rho}$ become relevant at two-loop order. We prove that $\nu_{\rho}$
is essentially the odd scalar curvature of an arbitrary torsion-free
connection that is compatible with both the anti-Poisson structure $E$ and
the density $\rho$. This extends a previous result for non-degenerate
antisymplectic manifolds to degenerate anti-Poisson manifolds that admit a
compatible two-form. 
\end{abstract}
\vfill
\begin{quote}
PACS number(s): 02.40.-k; 03.65.Ca; 04.60.Gw; 11.10.-z; 11.10.Ef; 11.15.Bt. 
\hrule width 5.cm \vskip 2.mm \noindent 
$^{1}${\small E-mail:~{\tt batalin@lpi.ru}} \hspace{10mm}
$^{2}${\small E-mail:~{\tt bering@physics.muni.cz}} \\
\end{quote}

\newpage

\section{Introduction}
\label{secintro}

\noi
The main purpose of this Letter is to report on new geometric insights into
the field-antifield formalism. In general, the field-antifield formalism
\cite{bv81,bv83,bv84} is a recipe for constructing Feynman rules for Lagrangian
field theories with gauge symmetries. The field-antifield formalism is in
principle able to handle the most general gauge algebra, \ie open gauge
algebras of reducible type. The input is usually a local relativistic field
theory, formulated via a classical action principle in a geometric
configuration space. In the field-antifield scheme, the original field
variables are extended with various stages of ghosts, antighosts and Lagrange
multipliers --- all of which are then further extended with corresponding
antifields; the gauge symmetries are encoded in a nilpotent Fermionic BRST
symmetry \cite{brs74,tyutin75}; and the original action is deformed into a
BRST-invariant master action, whose Hessian has the maximal allowed rank. The
full quantum master action 
\beq
W~=~S+\sum_{n=1}^{\infty}\hbar^{n} M_{n}
\eeq  
is determined recursively order by order in \mb{\hbar} from a consistent set of
quantum master equations
\bea
(S,S)&=&0~, \label{cme} \\
(M_{1},S)&=&i(\Delta_{\rho}S)~, \label{qme1} \\
(M_{2},S)&=&i(\Delta_{\rho}M_{1})+\nurho-\Hf(M_{1},M_{1})~, \label{qme2} \\
(M_{n},S)&=&i(\Delta_{\rho}M_{n-1})-\Hf\sum_{r=1}^{n-1} (M_{r},M_{n-r})
~,~~~~~~n\geq3~. \label{qme3}
\eea
Here \mb{(\cdot,\cdot)} is the antibracket (or anti-Poisson structure), 
\mb{\Delta_{\rho}} is the odd Laplacian and \mb{\nurho} is an odd scalar, which
become relevant in perturbation theory at loop order \mb{0}, \mb{1}, and
\mb{2}, respectively. It has only recently been realized that the
field-antifield formalism can consistently accommodate a non-zero \mb{\nurho}
term, thereby providing a more flexible framework for field-antifield
quantization \cite{b06,b07,bb07}. 

\noi
The classical master equation \e{cme} is a generalization of Zinn-Justin's
equation \cite{zinnjustin75}, which allows to set up consistent renormalization
(if the field theory is renormalizable). If the theory is not anomalous at the
one-loop level, there will exist a local solution \mb{M_{1}} to the next
equation \e{qme1}, and so forth. Although the field-antifield formalism in its
basic form is only a formal scheme --- \ie particularly, it assumes that
results from finite dimensional analysis are directly applicable to field
theory, which has infinitely many degrees of freedom --- it has nevertheless
been successfully applied to a large variety of physical models. It has mainly
been used in a truncated form of the full set of quantum master eqs.\ 
\e{cme} -- \e{qme3}, where all the following quantities 
\beq
(S,S),~(\Delta_{\rho}S), ~\nurho, ~M_{1},~M_{2},~M_{3},\ldots, \label{trunc}
\eeq
are set identically equal to zero. One can for instance mention the AKSZ 
paradigm \cite{aksz97,r07} as a broad example that uses the truncated
field-antifield formalism \e{trunc} to quantize supersymmetric topological
field theories \cite{ikeda94,ss94,cf99,bm01}. Currently, very few scientific
works describe solutions with non-zero \mb{M_{n}}'s, primarily due to the
singular nature of the odd Laplacian \mb{\Deltarho} in field theory (again 
because of the infinitely many degrees of freedom). Nevertheless, it should be
fruitful to study generic solutions of the full quantum master equation. See
the original paper \cite{bv81} for an interesting solution with
\mb{M_{1}\neq 0}. {}Finally, it has in many cases been explicitly checked that
the field-antifield formalism produces the same result as the Hamiltonian
formulation \cite{ggt91,dfgh91,fdj93}. The formalism has also influenced work
in closed string field theory \cite{zwiebach93} and several branches of
mathematics. The geometry behind the field-antifield formalism was further
clarified in \Ref{schwarz93,bt93,kn93,hz94}.

\noi
In this Letter we shall only explicitly consider the case of finitely many
variables. Our main result concerns the odd scalar \mb{\nurho}, which is a
certain function of the anti-Poisson structure \mb{E^{AB}} and the density
\mb{\rho}, \cf \eq{nurho} below. It turns out that \mb{\nurho} has a geometric
interpretation as (minus \mb{1/8} times) the odd scalar curvature \mb{R} of any
connection \mb{\nabla} that satisfies three conditions; namely that \mb{\nabla}
is 1) anti-Poisson, 2) torsion-free and 3) \mb{\rho}-compatible. This is a
rather robust conclusion as we shall prove in this Letter that it even holds for
degenerate antibrackets. (Degenerate anti-Poisson structures appear naturally
from for instance the Dirac antibracket construction for antisymplectic
second-class constraints \cite{b07,bt93,bbd97,bbd06}.)

\section{Anti-Poisson structure \mb{E^{AB}}}
\label{secap}

\noi
An {\em anti-Poisson} structure is by definition a possibly degenerate
\mb{(2,0)} tensor field \mb{E^{AB}}  with upper indices that is Grassmann-odd
\beq
 \eps(E^{AB})~=~\eps_{A}+\eps_{B}+1~,\label{upperodd}
\eeq
that is skewsymmetric 
\beq
E^{AB}~=~-(-1)^{(\eps_{A}+1)(\eps_{B}+1)}E^{BA}~,\label{upperskewsym}
\eeq
and that satisfies the Jacobi identity
\beq
\sum_{{\rm cycl.}~A,B,C}(-1)^{(\eps_{A}+1)(\eps_{C}+1)}
E^{AD} (\lpa{D}E^{BC}) ~=~0~.\label{abjacid}
\eeq

\section{Compatible two-form \mb{E_{AB}}}
\label{seccomptwoform}

\noi
In general, an anti-Poisson manifold could have singular points where the rank
of \mb{E^{AB}} jumps, and it is necessary to impose a regularity criterion to 
proceed. We shall here assume that the anti-Poisson structure \mb{E^{AB}}
admits a compatible two-form field \mb{E_{AB}}, \ie that there exists a
two-form field \mb{E_{AB}} with lower indices that is Grassmann-odd
\beq
 \eps(E_{AB})~=~\eps_{A}+\eps_{B}+1~,\label{lowerodd}
\eeq
that is skewsymmetric 
\beq
E_{AB}~=~-(-1)^{\eps_{A}\eps_{B}}E_{BA}~,\label{lowerskewsym}
\eeq
and that is {\em compatible} with the anti-Poisson structure in the sense that
\bea
 E^{AB}E_{BC}E^{CD}&=&E^{AD}~,\label{uppertripleeee} \\
 E_{AB}E^{BC}E_{CD}&=&E_{AD}~.\label{lowertripleeee}
\eea
This is a relatively mild requirement, which is always automatically satisfied
for a Dirac antibracket on antisymplectic manifolds with antisymplectic
second-class constraints \cite{b07,bt93,bbd97,bbd06}. Note that the two-form
\mb{E_{AB}} is neither unique nor necessarily closed. One can define a
\mb{(1,1)} tensor field as
\beq
P^{A}{}_{C}~\equiv~ E^{AB}E_{BC}~,
\eeq
or equivalently,
\beq
P_{A}{}^{C}~\equiv~ E_{AB}E^{BC} ~=~ (-1)^{\eps_{A}(\eps_{C}+1)}P^{C}{}_{A}~.
\eeq
It then follows from either of the compatibility relations
\es{uppertripleeee}{lowertripleeee} that \mb{P^{A}{}_{B}} is an idempotent
\beq
P^{A}{}_{B} P^{B}{}_{C}~=~P^{A}{}_{C}~.
\eeq

\section{The \mb{\DeltaE} Operator}
\label{secdeltae}

\noi
An anti-Poisson structure with a compatible two-form field \mb{E_{AB}} gives
rise to a Grassmann-odd, second-order \mb{\DeltaE} operator that takes
semidensities to semidensities. It is defined in arbitrary coordinates as
\cite{b07}
\beq
\DeltaE~\equiv~\Deltaone
+\frac{\nu^{(1)}}{8}-\frac{\nu^{(2)}}{8}-\frac{\nu^{(3)}}{24}
+\frac{\nu^{(4)}}{24}+\frac{\nu^{(5)}}{12}~,
\label{deltae}
\eeq
where \mb{\Deltaone} is the odd Laplacian 
\beq
\Deltarho~\equiv~\frac{(-1)^{\eps_{A}}}{2\rho}\lpa{A}\rho E^{AB}\lpa{B}
~,\label{deltarho} 
\eeq
with \mb{\rho=1}, and where
\bea
\nu^{(1)}&\equiv&
(-1)^{\eps_{A}}(\lpa{B}\lpa{A}E^{AB})~,\label{nu1} \\
\nu^{(2)}&\equiv& (-1)^{\eps_{A}\eps_{C}}(\lpa{D}E^{AB})E_{BC}
(\lpa{A}E^{CD})~,\label{nu2} \\
\nu^{(3)}&\equiv&(-1)^{\eps_{B}}(\lpa{A}E_{BC})
E^{CD}(\lpa{D}E^{BA})~,\label{nu3} \\
\nu^{(4)}&\equiv&(-1)^{\eps_{B}}(\lpa{A}E_{BC})
E^{CD}(\lpa{D}E^{BF})P_{F}{}^{A}~,\label{nu4} \\
\nu^{(5)}&\equiv&(-1)^{\eps_{A}\eps_{C}}(\lpa{D}E^{AB})E_{BC}
(\lpa{A}E^{CF})P_{F}{}^{D} \cr
&=&(-1)^{(\eps_{A}+1)\eps_{B}} E^{AD}(\lpa{D}
E^{BC})(\lpa{C}E_{AF})P^{F}{}_{B}~.\label{nu5} 
\eea
It is shown in \Ref{b07} that the \mb{\DeltaE} operator defined in \eq{deltae} 
does not depend on the choice of local coordinates, it does not depend on the
choice of compatible two-form field \mb{E_{AB}}, and it does map semidensities 
into semidensities. Moreover, the Jacobi identity \e{abjacid} precisely ensures
that \mb{\DeltaE} is nilpotent
\beq
\Delta_{E}^{2}~=~\Hf [\DeltaE,\DeltaE]~=~0~.\label{deltaenilp}
\eeq
Earlier works on the \mb{\DeltaE} operator include
\Ref{b06,bbd06,k99,kv02,k02,k04}.

\section{The \mb{\Delta} Operator}
\label{secdelta}

\noi
Classically, the field-antifield formalism is governed by the anti-Poisson
structure \mb{E^{AB}}, or equivalently, the antibracket
\beq
(f,g)~\equiv~(f\rpa{A})E^{AB}(\lpa{B}g)
~=~-(-1)^{(\eps_{f}+1)(\eps_{g}+1)}(g,f)~.
\label{antibracket}
\eeq
Quantum mechanically, the field-antifield recipe instructs one to choose an
arbitrary path integral measure \mb{\rho}, and to use it to build a nilpotent, 
Grassmann-odd, second-order \mb{\Delta} operator that takes scalar functions
into scalar functions. It is natural to build the \mb{\Delta} operator by
conjugating the \mb{\DeltaE} operator \e{deltae} with appropriate square roots
of the density \mb{\rho} as follows:
\beq
\Delta~\equiv~\frac{1}{\sqrt{\rho}}\DeltaE\sqrt{\rho}~. \label{deltadeltaerho}
\eeq
In this way the \mb{\Delta} operator trivially inherits the nilpotency property
from the \mb{\DeltaE} operator,
\beq
\Delta^{2}~=~\frac{1}{\sqrt{\rho}}\Delta_{E}^{2}\sqrt{\rho}~=~0~.
\label{deltanilp}
\eeq
In physical applications the nilpotency \e{deltanilp} of \mb{\Delta} is
important for the underlying BRST symmetry of the theory.

\section{The Odd Scalar \mb{\nurho}}
\label{secnurho}

\noi
The odd scalar function \mb{\nurho} is defined as
\beq
\nurho~\equiv~ (\Delta 1)~=~\frac{1}{\sqrt{\rho}}(\DeltaE\sqrt{\rho})
~=~ \nu_{\rho}^{(0)}+\frac{\nu^{(1)}}{8}-\frac{\nu^{(2)}}{8}
-\frac{\nu^{(3)}}{24}+\frac{\nu^{(4)}}{24}+\frac{\nu^{(5)}}{12}~,
\label{nurho}
\eeq
where \mb{\nu^{(1)}}, \mb{\nu^{(2)}}, \mb{\nu^{(3)}}, \mb{\nu^{(4)}},
\mb{\nu^{(5)}} are given in eqs.\ \e{nu1}--\e{nu5}, and the quantity
\mb{\nu_{\rho}^{(0)}} is given as
\beq
\nu_{\rho}^{(0)}~\equiv~\frac{1}{\sqrt{\rho}}(\Deltaone\sqrt{\rho})~.
\label{nurho0}
\eeq 
The second-order \mb{\Delta} operator \e{deltadeltaerho} decomposes as
\beq
 \Delta~=~\Deltarho+\nurho~, \label{deltadeltarhonurho}
\eeq
where \mb{\Deltarho} is the odd Laplacian \e{deltarho}. 
The nilpotency of \mb{\Delta} implies that
\bea
 \Delta_{\rho}^{2}&=&(\nurho \, , \, \cdot \, )~,\label{absnil} \\
 (\Deltarho\nurho)&=&0~. \label{deltaannihilatesnu}
\eea
The possibility of a non-trivial \mb{\nurho} has only recently been observed,
\cf \Ref{b06,b07,bb07}. In the past, the odd scalar term \mb{\nurho} was not
present due to a certain compatibility relation between \mb{E} and \mb{\rho},
which was unnecessarily imposed, and which (using our new terminology) made
\mb{\nurho} vanish. In terms of the quantum master equation
\beq
\Delta e^{\Ih W}~=~0~, \label{qme}
\eeq 
the odd scalar \mb{\nurho} enters at the two-loop order
\mb{{\cal O}(\hbar^{2})}
\beq
\Hf(W,W)~=~i\hbar\Deltarho W+\hbar^{2}\nurho~, \label{mqme} 
\eeq
which in turn leads to the set of eqs.\ \e{cme} -- \e{qme3}.

\section{Connection}
\label{secconn}

\noi
In the next two Sections~\ref{secconn} and \ref{seccurv} we will briefly state
our sign conventions and definitions for the covariant derivative and the
curvature in the presence of Fermionic degrees of freedom. A more complete
treatment can be found in \Ref{bb07,b97}. Other references include
\Ref{lavrov04}. Our convention for the left covariant derivative
\mb{(\nabla_{A}X)^{B}} of a left vector field \mb{X^{A}} is \cite{b97}
\beq
(\nabla_{A}X)^{B}~\equiv~(\lpa{A}X^{B})
+(-1)^{\eps_{X}(\eps_{B}+\eps_{C})}\Gamma_{A}{}^{B}{}_{C}X^{C}~,~~~~~~~
\eps(X^{A})~=~\eps_{X}+\eps_{A}~.\label{nabladef}
\eeq 
A connection \mb{\Gamma_{A}{}^{B}{}_{C}} is called {\em anti-Poisson} if it
preserves the anti-Poisson structure \mb{E^{AB}}, \ie
\beq
0~=~(\nabla_{A}E)^{BC}
~\equiv~(\lpa{A}E^{BC})+\left(\Gamma_{A}{}^{B}{}_{D}E^{DC}
-(-1)^{(\eps_{B}+1)(\eps_{C}+1)}(B\leftrightarrow C)\right)~.
\label{connantipoisson}
\eeq
It is useful to define a reordered Christoffel symbol \mb{\Gamma^{A}{}_{BC}} 
as  
\beq
\Gamma^{A}{}_{BC}~\equiv~(-1)^{\eps_{A}\eps_{B}}\Gamma_{B}{}^{A}{}_{C}~.
\label{alternativechristoffel}
\eeq
A {\em torsion-free} connection \mb{\Gamma^{A}{}_{BC}} has the following
symmetry in the lower indices:
\beq
\Gamma^{A}{}_{BC}~=~-(-1)^{(\eps_{B}+1)(\eps_{C}+1)}\Gamma^{A}{}_{CB}~.
\label{torsionfree}
\eeq
A connection \mb{\Gamma^{A}{}_{BC}} is called \mb{\rho}-{\em compatible} if 
\beq
\Gamma^{B}{}_{BA}~=~(\ln\rho~\rpa{A})~.
\label{connrho}
\eeq
There are in principle two definitions for the divergence \mb{{\rm div}X} of a
Bosonic vector field \mb{X} with \mb{\eps_{X}\!=\!0}. The first divergence
definition depends on the density \mb{\rho}
\beq
{\rm div}_{\rho}^{}X~\equiv~\frac{(-1)^{\eps_{A}}}{\rho} 
\lpa{A}(\rho X^{A})~, \label{divrho}
\eeq
while the second definition depends on the connection \mb{\nabla}
\beq
{\rm div}_{\nabla}^{}X~\equiv~\str(\nabla X)~\equiv~
(-1)^{\eps_{A}}(\nabla_{A}X)^{A}
~=~((-1)^{\eps_{A}}\lpa{A}+\Gamma^{B}{}_{BA})X^{A}~.\label{divnabla}
\eeq
The \mb{\rho}-compatibility condition \e{connrho} precisely ensures that the
two definitions \es{divrho}{divnabla} coincide, and hence that there is a
unique notion of volume \cite{yks02}. We shall only consider torsion-free
connections \mb{\nabla} that are anti-Poisson and \mb{\rho}-compatible, \ie
connections that satisfy the above three conditions \e{connantipoisson},
\es{torsionfree}{connrho}. Then the odd Laplacian \mb{\Deltarho} can be written
on a manifestly covariant form
\beq
 \Deltarho~=~\frac{(-1)^{\eps_{A}}}{2}\nabla_{A}E^{AB}\nabla_{B}
~=~\frac{(-1)^{\eps_{B}}}{2}E^{BA}\nabla_{A}\nabla_{B}~.\label{covdeltarho}
\eeq

\section{Curvature}
\label{seccurv}

\noi
The Riemann curvature tensor is
\beq
R^{A}{}_{BCD}~\equiv~(-1)^{\eps_{A}\eps_{B}}(\lpa{B}\Gamma^{A}{}_{CD})
+\Gamma^{A}{}_{BE}\Gamma^{E}{}_{CD}
-(-1)^{\eps_{B}\eps_{C}}(B\leftrightarrow C)~. \label{riemanncurvaturetensor}
\eeq
(Note that the ordering of indices on the Riemann curvature tensor is slightly
non-standard to minimize appearances of sign factors.) The Ricci tensor is
\beq
R_{AB}~\equiv~R^{C}{}_{CAB} 
~=~\frac{(-1)^{\eps_{C}}}{\rho}(\lpa{C}\rho\Gamma^{C}{}_{AB})
-(\lpa{A}\ln\rho~\rpa{B})-\Gamma_{A}{}^{C}{}_{D}\Gamma^{D}{}_{CB}
~=~-(-1)^{(\eps_{A}+1)(\eps_{B}+1)}R_{BA}~.\label{riccitensor} 
\eeq

\section{Odd Scalar Curvature}
\label{secosc}

\noi
The odd scalar curvature \mb{R} is defined as the Ricci tensor \mb{R_{AB}}
contracted with the anti-Poisson tensor \mb{E^{AB}},
\beq
R~\equiv~R_{AB}E^{BA}~=~E^{AB}R_{BA}~,~~~~~~~~~~~~\eps(R)~=~1~.\label{defosc}
\eeq
We now assert that the odd scalar curvature 
\beq
R~=~-8\nurho \label{rnurho}
\eeq
of an arbitrary connection \mb{\nabla} that is anti-Poisson, torsion-free and 
\mb{\rho}-compatible, is equal to (minus eight times) the odd scalar
\mb{\nurho}. In particular one sees that the odd scalar curvature \mb{R}
carries no information about the connection \mb{\nabla} used, and it depends
only on \mb{E} and \mb{\rho}. Equation \e{rnurho} was proven for the
non-degenerated case in \Ref{bb07}. The degenerated case is proven in
Appendix~\ref{appproof}. 

\vspace{0.8cm}

\noi
{\sc Acknowledgement:}~We would like to thank P.H.~Damgaard for discussions. 
K.B.\ thanks the Lebedev Physics Institute and the Niels Bohr Institute for
warm hospitality. The work of I.A.B.\ is supported by grants RFBR 05-01-00996, 
RFBR 05-02-17217 and LSS-4401.2006.2. The work of K.B.\ is supported by the
Ministry of Education of the Czech Republic under the project MSM 0021622409.

\appendix

\section{Proof of the Main Eq.\ \e{rnurho}}
\label{appproof}

Equation (C.9) in \Ref{bb07} yields that the odd scalar curvature \mb{R}
can be written as
\beq
R~=~-8\nu_{\rho}^{(0)}-\nu^{(1)}-\Hf R_{I}~,\label{earlyrnurho}
\eeq
where \mb{\nu_{\rho}^{(0)}}, \mb{\nu^{(1)}} and \mb{R_{I}} are defined in eqs.\
\e{nurho0}, \es{nu1}{r1}, respectively. Since the expression \e{r1} below for 
\mb{R_{I}} only depends on the torsion-free part of the connection, one does in
principle not need the torsion-free condition \e{torsionfree} from now on.
The heart of the proof consists of the following ten ``one-line calculations'':
\bea
R_{I}&\equiv&\Gamma^{A}{}_{BC}(E^{CB}\rpa{A})
~=~\Gamma^{A}{}_{BC}((E^{CD}E_{DF}E^{FB})\rpa{A})
~=~2R_{II}+R_{III}~,\label{r1} \\
R_{II}&\equiv&\Gamma^{A}{}_{BC}P^{C}{}_{D}(E^{DB}\rpa{A})
~=~-R_{IV}-\nu^{(2)}~,\label{r2} \\
R_{III}&\equiv&(-1)^{\eps_{A}(\eps_{C}+1)}
\Gamma_{F}{}^{A}{}_{B}E^{BC}(\lpa{A}E_{CD})E^{DF}
~=~2R_{III}+R_{V}~,\label{r3} \\
R_{IV}&\equiv&\Gamma^{A}{}_{BC}E^{CD}(\lpa{D}E^{BF})E_{FA}
~=~R_{VI}-R_{IV}~,\label{r4} \\
R_{V}&\equiv&(-1)^{\eps_{A}\eps_{C}}
\Gamma_{F}{}^{A}{}_{B}P^{B}{}_{C}(\lpa{A}E^{CD})P_{D}{}^{F}
~=~R_{VII}-\nu^{(5)}~,\label{r5} \\
R_{VI}&\equiv&\Gamma^{A}{}_{BC}(E^{CB}\rpa{D})P^{D}{}_{A}
~=~2R_{VIII}+R_{IX}~,\label{r6} \\
R_{VII}&\equiv&(-1)^{(\eps_{A}+1)(\eps_{C}+1)}E_{AB}
\Gamma^{B}{}_{CD}E^{DF}(\lpa{F}E^{AG})P_{G}{}^{C}
~=~R_{IV}-R_{VIII}~,\label{r7} \\
R_{VIII}&\equiv&\Gamma^{A}{}_{BC}P^{C}{}_{D}(E^{DB}\rpa{F})P^{F}{}_{A}
~=~-R_{IV}-\nu^{(5)}~,\label{r8} \\
R_{IX}&\equiv&(-1)^{\eps_{A}(\eps_{C}+1)}
\Gamma_{G}{}^{A}{}_{B}E^{BC}P_{A}{}^{D}(\lpa{D}E_{CF})E^{FG}
~=~-R_{X}-\nu^{(4)}~,\label{r9} \\
R_{X}&\equiv&(-1)^{\eps_{A}}
\Gamma_{F}{}^{A}{}_{B}E^{BC}(\lpa{C}E_{AD})E^{DF}
~=~-R_{III}-\nu^{(3)}~.\label{r10}
\eea
Here we have used the upper compatibility relation \e{uppertripleeee} for the 
two-form \mb{E_{AB}} in the second equality of eqs.\ \e{r1}, \e{r6}, \e{r7},
\es{r8}{r9}; the lower compatibility relation \e{lowertripleeee} for the
two-form \mb{E_{AB}} in the second equality of \eq{r3}; the anti-Poisson
property \e{connantipoisson} for the connection \mb{\nabla} in the second
equality of eqs.\ \e{r2}, \e{r5}, \e{r8}, \es{r9}{r10}; and the Jacobi identity
\e{abjacid} in the second equality of \eqs{r4}{r7}. {}From these ten relations
\e{r1}--\e{r10}, the quantity \mb{R_{III}} can be determined as follows:
\bea
-R_{III}&=&R_{V}~=~R_{VII}-\nu^{(5)}
~=~(R_{IV}-R_{VIII})+(R_{IV}+R_{VIII})~=~2R_{IV} \cr 
&=&R_{VI}~=2R_{VIII}+R_{IX}
~=~-2(R_{IV}+\nu^{(5)})+(R_{III}+\nu^{(3)}-\nu^{(4)}) \cr
&=&2R_{III}+(\nu^{(3)}-\nu^{(4)}-2\nu^{(5)})~,
\eea
so that
\beq
  R_{III}~=~\frac{1}{3}(-\nu^{(3)}+\nu^{(4)}+2\nu^{(5)})~.\label{r3solved}
\eeq
Next, \mb{R_{I}} can be expressed in terms of \mb{R_{III}}:
\beq
\Hf R_{I}~=~R_{II}+\Hf R_{III}~=~-(R_{IV}+\nu^{(2)})+\Hf R_{III}
~=~R_{III}-\nu^{(2)}~.\label{r1asr3}
\eeq
Inserting \eqs{r3solved}{r1asr3} into \eq{earlyrnurho} yields the main
\eq{rnurho}:
\beq
R~=~-8\nu_{\rho}^{(0)}-\nu^{(1)}-\Hf R_{I} 
~=~-8\nu_{\rho}^{(0)}-\nu^{(1)}+\nu^{(2)}
+\frac{1}{3}(\nu^{(3)}-\nu^{(4)}-2\nu^{(5)})~=~-8\nurho~.
\eeq

\end{document}